\documentclass[journal=jacsat,manuscript=article]{achemso}
\usepackage{chemformula} 
\usepackage[T1]{fontenc} 
\usepackage{amsmath,amssymb,amsbsy}
\usepackage{subcaption}
\usepackage{wrapfig}
\usepackage{gensymb}
\usepackage{tabu}

\author{Tara M. Boland}
\affiliation[Arizona State University]
{School for Engineering of Matter, Transport and Energy, Arizona State University, Tempe, AZ, USA}

\author{Peter Rez}
\affiliation[Arizona State University]
{Department of Physics, Arizona State University, Tempe, AZ, USA}

\author{Peter A. Crozier}
\affiliation[Arizona State University]
{School for Engineering of Matter, Transport and Energy, Arizona State University, Tempe, AZ, USA}

\author{Arunima K. Singh}
\email{arunimasingh@asu.edu}
\affiliation[Arizona State University]
{Department of Physics, Arizona State University, Tempe, AZ, USA}

\title{Impact of Aliovalent Alkaline-Earth Metal Solutes on Ceria Grain Boundaries: A Density Functional Theory Study}

\abbreviations{GB: Grain boundary}
\abbreviations{GB geometry: The 5 macroscopic degrees of freedom which define the GB which describes the crystallographic relationship between the 2 neighbouring grains. Also called GB character.}
\abbreviations{GB structure: The atomic-level features of a GB e.g. atomic positions, defect structure, chemistry.}

\newcommand{\stooo}{$\Sigma$3 (111)/[$\bar{1}$01]}
\newcommand{\stoot}{$\Sigma$3 (121)/[$\bar{1}$01]}
\newcommand{\egb}{$\Delta E_{\mathrm{GB}}$}

\newcommand{\gbe}{$\gamma _{\mathrm{GB}}$}

\newcommand{\egap}{$E_{\mathrm{gap}}$}

\keywords{Ceria, Grain-Boundaries, Aliovalent Dopants}

\begin{document}
\noindent
------------------------------------------------------------------------------------------------------------------------



\begin{abstract}

Ceria has proven to be an excellent ion-transport and ion-exchange material when used in polycrystalline form and with a high-concentration of aliovalent doped cations. Despite its widespread application, the impact of atomic-scale defects in this material are scarcely studied and poorly understood. In this article, using first-principles simulations, we provide a fundamental understanding of the atomic-structure, thermodynamic stability and electronic properties of undoped grain-boundaries (GBs) and alkaline-earth metal (AEM) doped GBs in ceria. Using density-functional theory simulations, with a GGA+U functional, we find the \stooo\ GB is thermodynamically more stable than the \stoot\ GB due to the larger atomic coherency in the \stooo\ GB plane. We dope the GBs with $\sim$20\% [M]$_{GB}$ (M=Be, Mg, Ca, Sr, and Ba) and find that the GB energies have a parabolic dependence on the size of solutes, the interfacial strain and the packing density of the GB. We see a stabilization of the GBs upon Ca, Sr and Ba doping whereas Be and Mg render them thermodynamically unstable. The electronic density of states reveal that no defect states are present in or above the band gap of the AEM doped ceria, which is highly conducive to maintain low electronic mobility in this ionic conductor. The electronic properties, unlike the thermodynamic stability, exhibit complex inter-dependence on the structure and chemistry of the host and the solutes. This work makes advances in the atomic-scale understanding of aliovalent cation doped ceria GBs serving as an anchor to future studies that can focus on understanding and improving ionic-transport. 

\end{abstract}

\section{Introduction}

Doped polycrystalline electroceramic oxides are an important class of materials in which point defects in the bulk and grain boundaries play a key role in regulating mechanical, optical, thermal, magnetic, catalytic and charge transport properties \cite{Sutton1995, Browning2004, Lee2012, Ye2014, Bowman2015, Lin2015, Bowman2017, Feng2017, Nafsin2017, Bowman2019}. The transport and ion exchange functionalities of electroceramics make them suitable for many technological applications including catalysts \cite{Trovarelli2002, Paier2013}, solid electrolytes and electrodes \cite{Singhal2003,wachsman2011,Wachsman2012}, gas separation membranes \cite{Sunarso2008, Zhang2012}, gas sensing systems \cite{Haile1998}, and memristors \cite{Schweiger2017}. Many of the relevant oxides have fluorite or perovskite structures where oxygen transport occurs via thermally activated vacancy hopping and electron transport takes place via polaron hopping. In these oxides, aliovalent cation doping can be employed to introduce oxygen vacancies, manipulate oxygen migration energies, and regulate the concentration of mobile electrons and holes. 

 Many applications employ polycrystalline solids where the overall properties of the material are significantly impacted by the presence of grain boundaries. For example, ionic conductivity is degraded by space charge effects which block oxygen transport across GBs \cite{Franceschetti1981, Maier1995, Guo2001, Lee2001, Guo2003, Tuller2006, Tuller2010} and it is commonly assumed that there is a high concentration of immobile positively charged oxygen vacancies at the GB core which repel mobile vacancies. Simple Mott-Schottky and Gouy-Chapman models have been developed to treat space charge effects \cite{Guo2003, Avila2006, Avila2009, Tuller2010, Tuller2011, Kim2016, Kim2016}. Although successful in many ways, the Mott-Schottky and Gouy-Chapman models are built on the so-called “dilute-solute” or “non-interacting” defect assumption. There is a growing body of experimental evidence \cite{Lei2002, Browning2004, Lee2012, Shirpour2012, An2013, Lee2013, Bowman2015, Lin2015, Diercks2016} and theoretical predictions \cite{An2013, Lee2013, Mebane2015} that confirm solute cation concentrations at GBs which exceed the range of validity of the dilute solute assumption ($\sim$ 1\%). These observations are not surprising considering the typical temperatures employed for ceramic processing. There is a strong driving force for solute segregation to reduce the overall system energy due to cation size mismatch, electrostatic forces (i.e. GB core charge neutralization), and/or reduction in the GB energy \cite{Lei2002, Shirpour2012, Aidhy2014, Mebane2015, Nafsin2017}. The ionic conductivity behavior of GBs with high solute concentration is substantially enhanced contradicting the predictions of the dilute-solute space charge models \cite{Bowman2017, Bowman2019}. The origin for the conductivity increase is not currently understood and requires a fundamental investigation of the role of solutes on the atomic structure and bonding at grain boundaries.  
 
In this article, using first-principles simulations, we provide a fundamental understanding of the atomic-structure, stability and electronic properties of pristine as well as aliovalent, alkaline-earth metal (AEM) doped GBs in CeO$_2$.  We show that a local doping with $\sim$20\% [M]$_{GB}$ (M=Be, Mg, Ca, Sr, and Ba) has a significant impact on the thermodynamic stability of the GBs. Using density-functional theory simulations with a GGA+U functional we examine the structure, thermodynamic stability and coordination of atoms at the GB interface for two of the more frequently observed grain-boundaries in Ca-doped ceria,~\cite{bowman2016, Bowman2017} the \stooo\ and \stoot\ GB. We show that a local doping with $\sim$20\% [M]$_{GB}$ (M=Be, Mg, Ca, Sr, and Ba) has a significant impact on the thermodynamic stability of the GBs. Element-projected and orbital-projected density of states show that no defect states are present in or above the band gap of the AEM doped ceria, which is conducive to maintaining lower electronic mobilities that is necessary for good ionic transport. In addition, we find that the band gap of ceria can be modulated by up to 0.3 eV by selecting different AEM dopants at the ceria GB.    

    \begin{figure}[tp]
        \centering
        \includegraphics[width=1\linewidth]{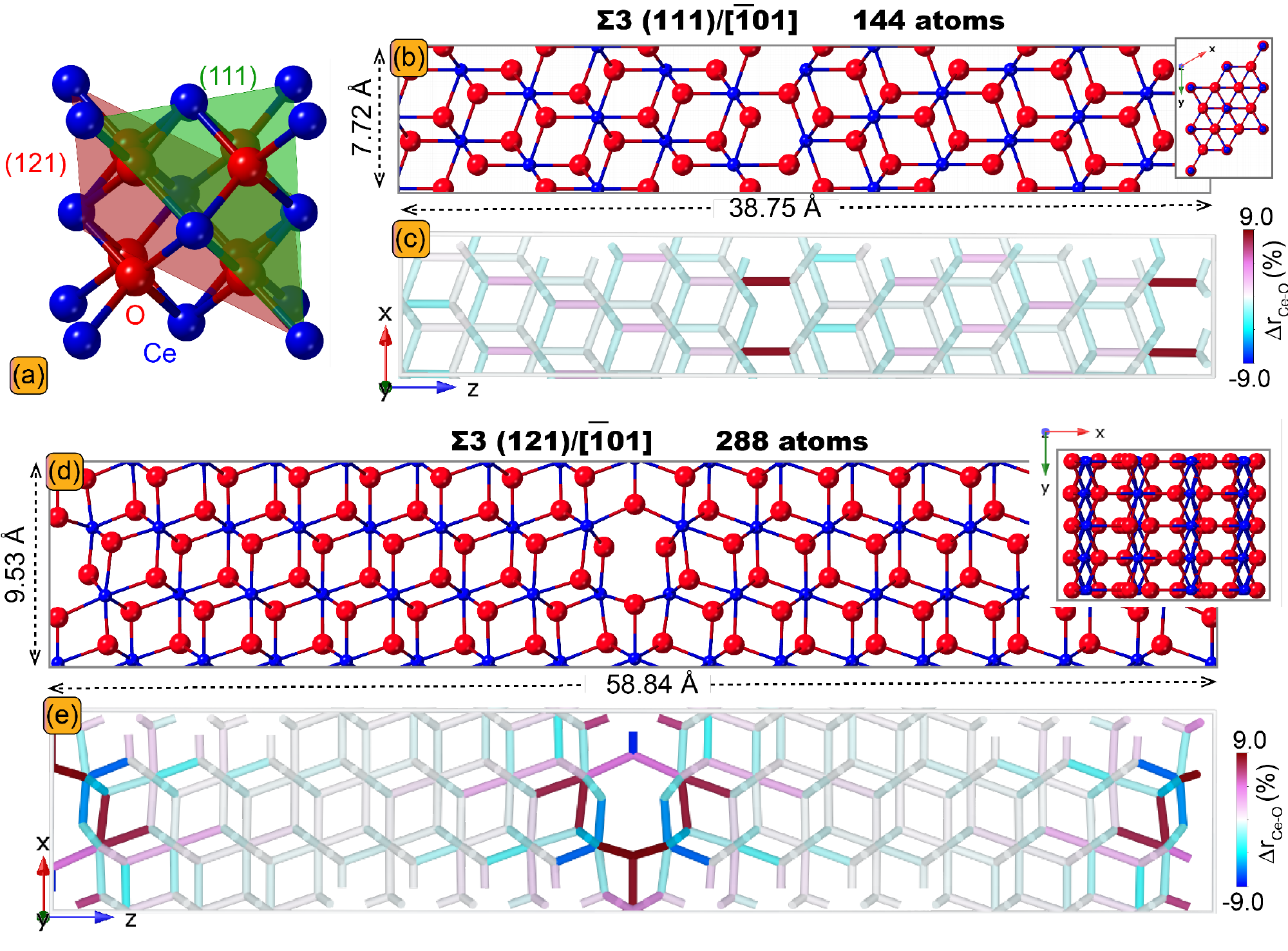}
        \caption{(a) Structure of the conventional fluorite CeO$_2$ unit cell. Its (121) and (111) lattice planes are shown as shaded planes. These indicate the interface planes for the GBs. (b) The xz-plane of the \stooo\ GB supercell. Inset shows the GB structure from the GB containing plane, i.e., the x-y plane and the GB is located in the center of the plane. (c) The percent deviation of the bonds in the \stooo\ GB supercell from the average Ce-O bond length of the GB supercell, $\Delta{r_{\mathrm{Ce-O}}}$, is shown as a color map. Blue indicates compressed bonds while red indicates tensile bonds. (d) The xz-plane of the \stoot\ GB supercell. Inset shows the structure from the x-y plane. (e) $\Delta{r_{\mathrm{Ce-O}}}$ is shown as a color map.}
        \label{fig:Figure1}
    \end{figure}
    
\section{Computational Methods}
 
    All simulations are based on density functional theory (DFT) using the the projector augmented wave method\cite{Blochl1994,Kresse5} as implemented in the plane-wave code VASP.\cite{Kresse1,Kresse2,Kresse3,Kresse4} All simulations included spin-polarization and the generalized gradient approximation (GGA) with the Perdew-Burke-Ernzerhof (PBE)\cite{Perdew19,Perdew20} exchange correlation functional was used. In addition, the strong correlation effects of the Ce 4\emph{f} electrons were treated within GGA using the Hubbard U correction (GGA+U) formulated by Dudarev et. al.\cite{Dudarev1998} An on-site Coulomb interaction, $U_{eff}$ = 5 eV, was used for Ce, as determined by Dholabhai et al.\cite{Dholabhai2010a}, to provide a better fit with the experimental band gap (\egap), lattice parameter ($a_0$), and bulk modulus ($B_0$) compared to traditional GGA methods. For a 2x2x2 supercell of bulk ceria, we find that $E_{\mathrm{gap}}$\big[O(2\emph{p}) $\rightarrow$ Ce(4\emph{f})\big]= 2.0 eV, $a_0$=5.5 \AA, and $B_0$=180.59 GPa which is in reasonable agreement with the measured values of $E_{\mathrm{gap}}$\big[O(2\emph{p}) $\rightarrow$ Ce(4\emph{f})\big]=3 eV\cite{Gerward1993}, $a_0$=5.411 \AA \cite{Eyring}, and $B_0$=204-236 GPa.\cite{Nakajima1994,Gerward1993} The chosen value of $U_{\mathrm{eff}}$ correctly describes the localization of the 4$f$ electrons on the nearby Ce atoms--unlike traditional GGA which results in delocalized electrons on all cerium ions in the lattice.

    A plane wave cutoff energy of 400 eV was used for all cases except for the volume optimization of ceria, where it was set to 520 eV. This cutoff energy was sufficient to converge the forces\cite{Hellman1937} acting on each ion to 0.01 eV \AA/atom or better. A block Davidson\cite{Broyden1965} minimization algorithm was used to achieve a convergence in total energy per cell on the order of 0.001 eV or better. 
    
    Figure~\ref{fig:Figure1}a and  Figure~\ref{fig:Figure1}b show the \stooo\ and \stoot\ GBs, respectively. These were constructed from the conventional fluorite unit cell of CeO$_2$, Figure~\ref{fig:Figure1}a, using pymatgen\cite{Ong2013}--an open-source Python library for materials analysis. The large variation in the atomic structures of these GBs and the presence of high-quality experimental characterization of GBs in polycrystalline Ca-doped ceria\cite{bowman2016,Bowman2017} motivates the choice of these two GBs for this study. The undoped \stooo\ GB cell has optimized lattice vectors [7.72, 7.72, 38.75] \AA\ with 144 atoms and was converged with a 3x3x1 gamma-centered $k$-point grid. The undoped \stoot\ GB cell has optimized lattice vectors [9.53, 7.74, 54.84] \AA\ with 288 atoms and was converged with a 2x3x1 gamma-center $k$-point grid. A Gaussian smearing with a sigma value of 0.05 eV was employed. 

    During the initial construction of each GB supercell structure, the inter-GB spacing between respective grains (the z-axis separation) was set to maintain the same cation-anion bond distance across the interfaces as the grain interior. This was motivated by several studies suggesting that ceramic oxides relax to retain a bond length between ions that is similar to the grain interiors.\cite{Shibata2004,Shibata2002} To minimize the GB interactions between periodic images, the undoped GB cells were constructed from grains having  a \textbf{c} lattice vector two times the periodic repeat distance of the oriented cell, such that $\textbf{c}=2\textbf{a}_{hkl}$, where $hkl$ are the crystal directions associated with the (111) and (121) interfacial planes. These GB supercells are used for assessing the energy and electronic properties of the undoped and doped GBs. Note that the GB energy difference between the GB supercells constructed using grains with \textbf{c}=\textbf{a}$_{hkl}$ and \textbf{c}=2\textbf{a}$_{hkl}$ was only 8 meV/\AA$^2$ and 2 meV/\AA$^2$ for the \stooo\ and \stoot\ GB, respectively. 

    We ensured that the strain fields due to the AEM solutes decayed within the supercell as discussed in the following section. When doped with an AEM solute, an oxygen vacancy was introduced in the cell to maintain charge neutrality. Pseudopotentials for each AEM solute were chosen such that the total energy was a minimum, and to ensure convergence of the simulations. The O and Ce atoms have been described by $2s^22p^4$ and $5s^25p^66s^25d^14f^1$ valence electrons, respectively. The valence electrons for Be and Mg were described by $2s^2$ and $3s^2$ while Ca, Sr, and Ba used $3s^23p^64s^2$, $4s^24p^65s^2$, $5s^25p^66s^2$ valence electrons, respectively. All structures, the bulk ceria, the GB structures and the AEM-doped GB, were subject to full structure optimization. 

\section{Results and discussion}

\subsection{Grain Boundary Structure and Character}

Grain boundary notations represent its 5 macroscopic degrees of freedom, i.e., the four degrees specifying two directions and one specifying the angle.\cite{Randle} Besides these macroscopic specifications, atomic-level parameters like the number of coordination-deficient cation sites, the average cation-anion bond distance, and the GB induced lattice expansion can further elucidate the GB's structure-property relationship. A coordination-deficient cation site is a site which has fewer bonds than that of the host cation in the defect-free lattice. Thus for ceria-based compounds, a coordination-deficient cation site will have less than 8 nearest neighbor oxygen atoms. GB expansion, \gbe\ in \AA, is defined as the difference in the $z$-axis length between the relaxed GB supercell and the corresponding relaxed GB-free supercell divided by two. Hence, \gbe\ is a measure of the expansion of the pristine ceria's lattice vector that is perpendicular to the GB plane. 

Table \ref{tab:Table1} lists the aforementioned atomic-scale parameters and the misorientation angles of the GBs. The \stooo\ and \stoot\ are both high-angle coincident site lattice boundaries~\cite{bowman2016} with misorientation angles of 35.26 and 54.74$^{\circ}$, respectively. Interestingly, the equidistant (near cubic) polyhedral arrangement of the O ions around the Ce ions tend to remain intact at/near the GB core as can be seen in Figure~\ref{fig:Figure1}. This can be attributed to the large ionicity of the Ce-O bonds. In order to retain the polyhedral arrangement of the host lattice the \gbe\ is significant, 0.315 \AA\ for \stooo\ and 0.471 \AA\ \stoot\ GB, in agreement with experimentally measured values in similar systems.\cite{Shibata2004} We emphasize here that in stoichiometric ceria, the coordination-deficient vacancy sites are structural in origin. The charge neutrality of the compound is maintained for all simulations thus no other point defects were considered to be present at the GB.

The averaged Ce-O bond distance, $\bar{r}^{~GB}_{\mathrm{Ce-O}}$, in the \stooo\ and \stoot\ GB models are 2.379 \AA\ and 2.385 \AA, respectively. These average bond distances in the GBs are practically equal to the bond distances in bulk ceria, $r^{CeO_2}_{\mathrm{Ce-O}}$ = 2.380 \AA. The excellent agreement between $\bar{r}^{~GB}_{\mathrm{Ce-O}}$ and $r^{CeO_2}_{\mathrm{Ce-O}}$, however, does not imply that there are no distortions in the lattice upon incorporation of the GB. On the contrary, as shown in Figure~\ref{fig:Figure1}c and d, up to $\pm$9\% bond deviation, $\Delta{r^{~GB}_{\mathrm{Ce-O}}} = \frac{r^{~GB}_{\mathrm{Ce-O}} - r^{CeO_2}_{\mathrm{Ce-O}}}{r^{CeO_2}_{\mathrm{Ce-O}}}\times 100$, where $r^{~GB}_{\mathrm{Ce-O}}$ is the length of bonds in the GB structure, is observed. Both tensile and compressive strains are present in each GB lattice. The lattice distortions are predominant near the GB and diminish rapidly away from the GB. 

    \begin{table}
        \centering
        {\tabulinesep=1mm
        \begin{tabu}{|c|c|r|c|r|c|c|}
        \hline
           Interface-Plane & $\theta$ ($^{\circ}$) & sites & $\bar{r}^{~GB}_{\mathrm{Ce-O}}$ (\AA) & $\gamma_{\mathrm{GB}}$ (\AA) & \egb\ (eV/\AA$^2$) \\
        \hline
           \stooo\ & 35.26 & 4 & 2.379 & 0.315 & 0.058 (0.93) \\
        \hline
           \stoot\ & 54.74 & 4 & 2.385 & 0.471 & 0.093 (1.48) \\
        \hline
        \end{tabu}}
        \caption{The interface-plane notation, the misorientation angle, $\theta$ in $^{\circ}$, the total number of coordination deficient cation sites per GB, the average Ce-O bond distance for each GB structure, $\bar{r}^{~GB}_{\mathrm{Ce-O}}$ in \AA, the z-axis expansion of the GB supercell, $\gamma_{\mathrm{GB}}$ in \AA\ and the GB energy, \egb\ in eV/\AA$^2$ are listed for the two GBs studied in this work. \egb\ values listed in parenthesis are in J/m$^2$.}
        \label{tab:Table1}
    \end{table}
    
\subsection{Thermodynamic Stability of GBs and Solute Doped GBs} 

    \begin{figure}[t]
        \includegraphics[width=0.5\linewidth]{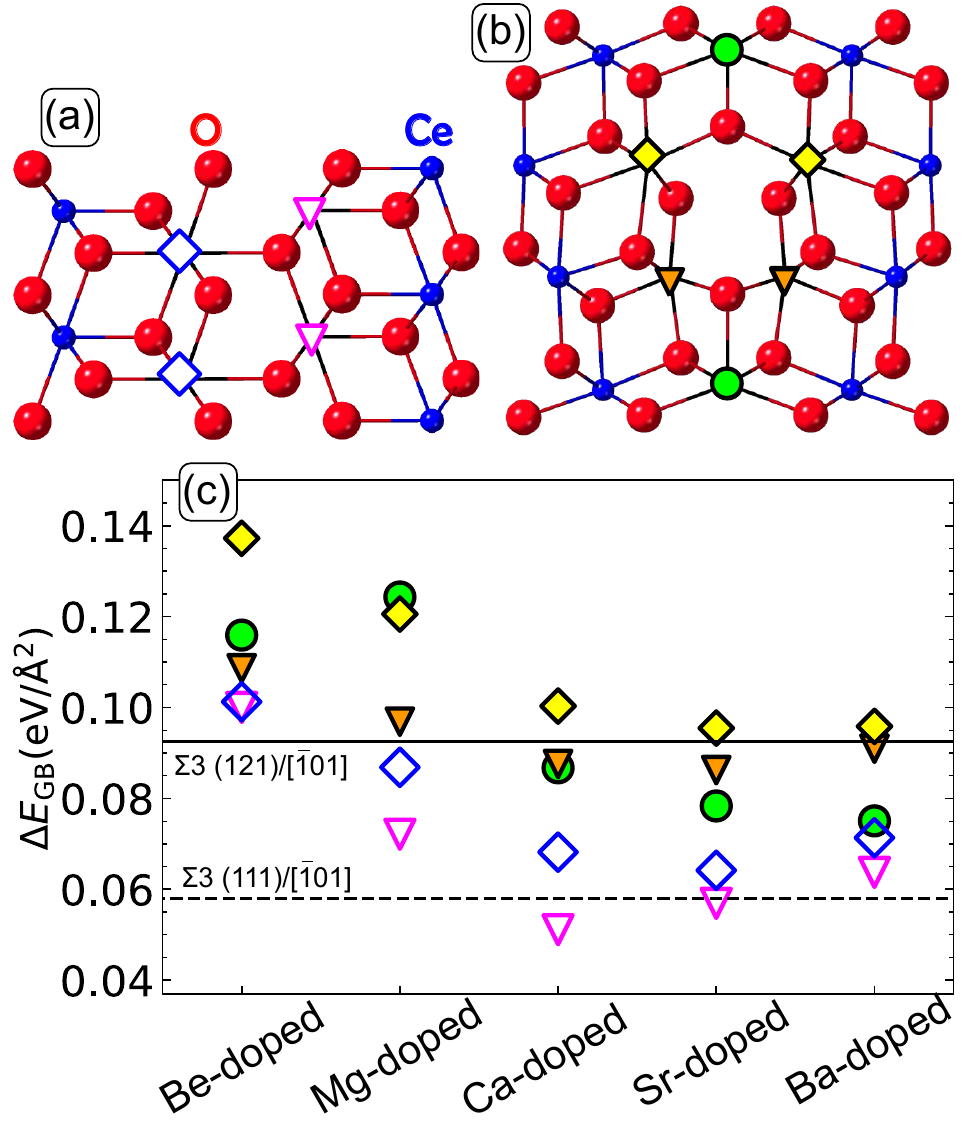}
        \caption{The a) \stooo\ and b) \stoot\ GB structures with all distinct solute sites indicated by a unique marker-color combination. All ions represented with a marker indicate a core GB cation site which was considered as potential substitutional site. c) \egb\ for the \stooo\ (open/dashed markers) and \stoot\ (filled/solid markers) GBs. Each marker corresponds to the \egb\ for each respective solute site depicted in the GB structure models.}
        \label{fig:Figure2}
    \end{figure}

In order to compare the stability of ceria in the presence of GBs' and dopants we compute the GB energy, \egb,

\begin{equation} \label{eq:GBEnergy}
\Delta E_{\mathrm{GB}}=\frac{ E_{\mathrm{GB}} - n_{\mathrm{CeO_2}} E_{\mathrm{CeO_2}}  - n_{\mathrm{MO}} E_{\mathrm{MO}} }{\mathrm{2A}}
\end{equation}

where $E_{\mathrm{GB}}$ is the total energy of the GB supercell with solute M, $E_{\mathrm{\chi}}$ is the energy of one formula unit of bulk $\chi$ where $\chi$ = CeO$_2$ or MO (see SI Table 1), $n_{\mathrm{\chi}}$ is the number of formula units of $\chi$ in the GB supercell, and A is the area of xy-plane i.e. the GB containing plane. \egb\ represents the area normalized excess energy of ceria due to the creation of the GB interface.  

As listed in Table \ref{tab:Table1}, the \egb\ of undoped \stooo\ GB is approximately half the value of the \stoot\ GB. This is not surprising since the \stooo\ GB has a high atomic coherency across the interface, see Figure \ref{fig:Figure1}b and SI Figure 1a. The continuity of the anion and cation sublattices is clearly preserved in the \stooo\ GB but the \stoot\ GB has a disruption in the cation sublattice, see Figure \ref{fig:Figure1}c and SI Figure 1b. Other ceramic oxides such as yttria stabilized zirconia, display similar dependence of the \egb\ on the coherency of atoms at the interface.~\cite{Shibata2002,Shibata2003,Shibata2004}

Figure~\ref{fig:Figure2}a and \ref{fig:Figure2}b mark the substitutional sites at the \stooo\ and \stoot\ GB of ceria, respectively, where we place the Be, Mg, Ca, Sr, and Ba solutes to assess their impact on the stability, structure and electronic properties of the lattice. Note that higher concentrations of dopants in ceria-based electrolytes have been reported at or near the GBs.\cite{Dholabhai2015,Bokov2018,bowman2016} By definition, for a cation site to be considered part of the GB core, the site must lie along/on either side of the GB mirror plane (see SI Figure 1). The \stooo\ GB has two distinct sites, a coordination-deficient site marked by magenta triangles and a fully-coordinated site marked by blue diamonds. The \stoot\ GB has three distinct sites, the coordination-deficient site marked by orange triangles and fully-coordinated sites marked by yellow diamonds and green circles. While the sites marked by green circles are fully-coordinated, they favor an asymmetric arrangement of the O-atoms around the site unlike the symmetric cubic arrangement in ceria. 

A local GB solute concentration of 25\% can be achieved for the \stooo\ and the \stoot\ GBs by sequentially considering one core GB site (indicated by the markers in Figure~\ref{fig:Figure2}a and \ref{fig:Figure2}b within one region of the GB core for doping. A region within the GB core is assumed to have a 2 \AA\ width perpendicular to the GB plane which originates at the cation mirror plane and extends towards the bulk. 
A total of 25 configurations of solutes were thus studied in this work. The large number of atoms in the simulation cell and the rapidly increasing number of configurations prohibit a comprehensive study of other solute concentrations.

Figure~\ref{fig:Figure2}c shows that the \egb\ is greater for the \stoot\ GB than the \stooo\ GB. Furthermore, for each substitutional site, \egb\ has a parabolic dependence on the solute cation's ionic radius. The site-dependence of the \egb\ of the \stooo\ GB is low in comparison to that of the \stoot\ GB. This can be understood by examining the net bond strain at the dopant sites of the GBs shown in Figure~\ref{fig:Figure1}c and \ref{fig:Figure1}e. In the \stoot\ GB, the three distinct solute sites have markedly different net bond strain illustrated by the variation in color in Figure~\ref{fig:Figure1}c, \ref{fig:Figure1}e, and SI Figure 10-11. The net tensile to compressive bond strain ratio is highest in the green-site, intermediate in the yellow site and lowest in the orange site. In comparison, the blue and magenta sites in the \stooo\ GB have a similar, and in fact much smaller, net bond strains shown in Figure~\ref{fig:Figure1}c.  

For both the GBs, the coordination-deficient cation sites (magenta and orange triangles) are among the lowest energy sites. For the \stoot\ GB, the fully-coordinated sites marked by the green circle also have low \egb, especially for the heavier solute cations. These three low \egb\ sites are also the most strained sites in the GBs. The blue sites in the \stooo\ GB and the yellow sites in the \stoot\ GB have largest energies displaying a barrier for doping and preference for the Ce-atoms to remain in a site that has coordination and bond-length similar to that of the grain interior. Similar trends in four symmetric tilt GBs have been observed for yttria-stablized zirconia.\cite{Shibata2002,Shibata2004} 

It is noteworthy that the addition of Be and Mg make the GBs consistently more unstable across all sites. Apart from the large mismatch in the ionic radii of Ce ($R_{i}$ = 0.97 \AA) with that of Be ($R_{i}$ = 0.27 \AA) and Mg ($R_{i}$ = 0.57 \AA)\cite{BioNum}, the nature of bonding in the native oxides of Mg and Be also dictates the stability of the GB. Unlike the octahedral coordination predominant in Ca ($R_{i}$ = 1.12 \AA), Sr ($R_{i}$ = 1.42 \AA) and Ba ($R_{i}$ = 1.26 \AA) oxides, Be and Mg oxides display a tetrahedral bonding, see SI Table 1. The Be and Mg dopants relax into interstitial sites to attain this 4-fold coordination where possible, for example in some of the coordination-deficient sites. The relaxed structure of all solutes configurations is presented in the SI Figure 10-11. Since the Ca dopants have the lowest mismatch in the ionic radii with the host Ce atoms and also more closely match the cubic coordination of the host cation, these solutes render the GB most stable in comparison to the other solutes. 

Overall, the \egb\ critical point appears to be modulated by three primary factors, (a) the local atomic environment of the solute site, (b) the solute size and (c) the coordination of the solute in its native oxide. The relative difference in \egb\ between the GBs may be due to the GB packing density. Furthermore, it is evident that the \egb\ can be more easily modulated by varying the solute type and is much less weakly modulated by the substitutional site. Additionally, GB doping strategies attempting to smooth out the potential energy landscape across GBs should focus on Ca or Sr solutes since out of the five solute sites explored the lowest GB energy is achieved for the Ca and Sr solutes.
       
\subsection{Electronic Structure of AEM Doped Ceria}
    Aliovalent solutes are often used to increase the number of charge carriers in ceramic oxides.\cite{Andersson2006} But they can also introduce localized defect states and/or bands above the band gap activating electronic conduction mechanisms such as polaron hopping.\cite{Bishop2011,Figueiredo2013} This can be detrimental to the ionic conductivities. In this section, we show that the AEM solutes can deactivate these potentially detrimental electronic conduction mechanisms. In this context, we find that AEM solutes do not introduce any defect states above the valence band or in the band gap as discussed below.

    \begin{figure}[ht]
        \centering
        \includegraphics[width=1\linewidth]{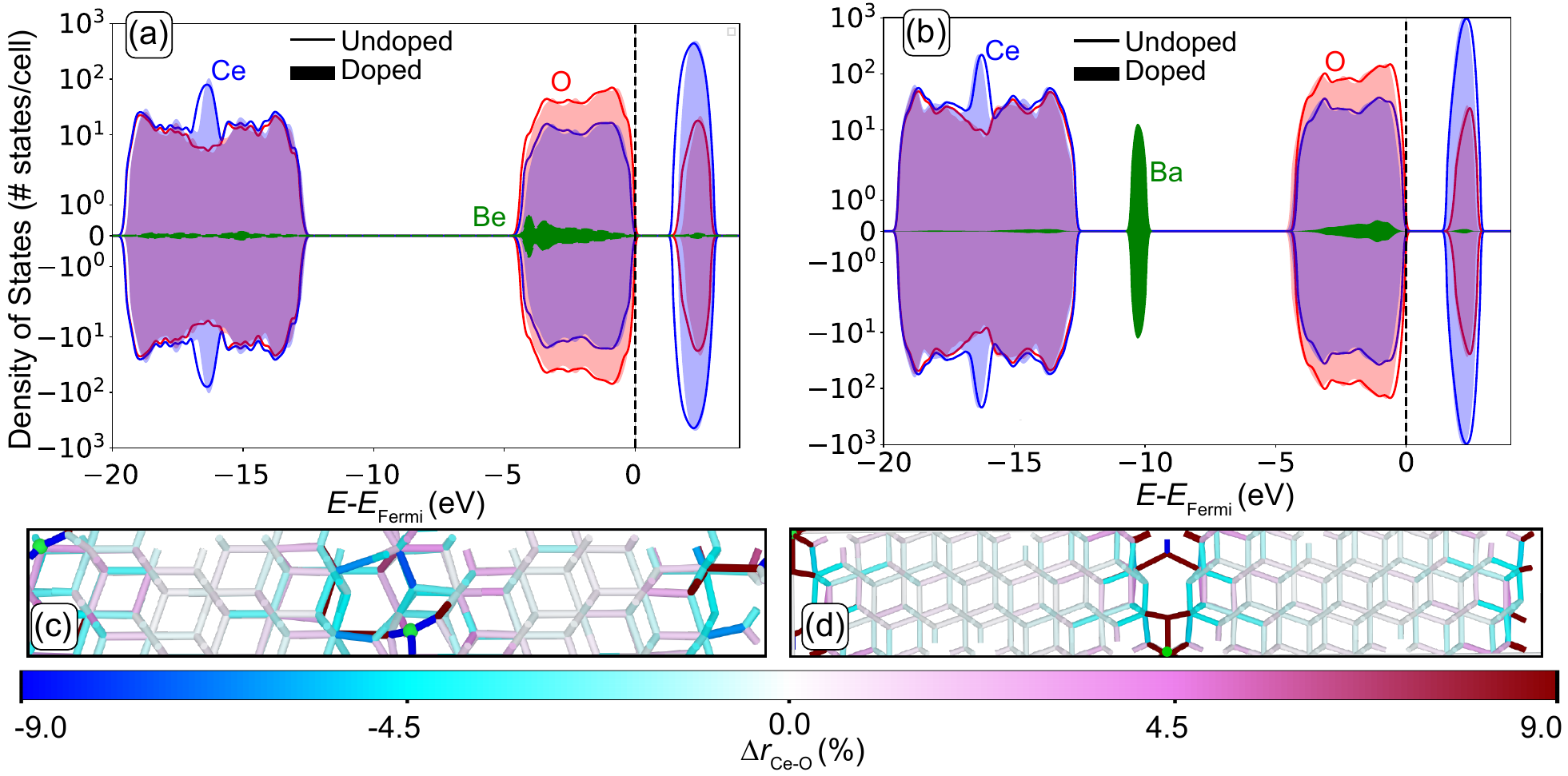}
        \caption{The element-projected DOS for the a) undoped (solid lines) and Be-doped (shaded regions) for the \stooo\ GBs, and b) undoped (solid lines) and Ba-doped (shaded regions) for the \stoot\ GBs. The $\Delta{r^{~GB}_{\mathrm{Ce-O}}}$ of the Be-doped \stooo\ GB is shown in (c) and that of the Ba-doped \stoot\ GB is shown in (d). Blue indicates compressed bonds while red indicates tensile bonds. The DOS is shifted such that the top of the valence band is at 0 eV.}
        \label{fig:Figure3}
    \end{figure}
    
Figure~\ref{fig:Figure3}a and b show the element-projected density of states (DOS) for the supercells with the \stooo\ and \stoot\ GB, respectively. The solid lines show the DOS for the undoped GB and the shaded regions mark the DOS for the solute-doped GB. From the DOS, it is clear that incorporation of Be and Ba solutes at the GB core does not result in defect states above the band gap or within it. Similarly, we find that none of the solutes impart defect states, see SI Figure 5-9.
    
In the undoped GB, the states at the conduction band maxima (CBM) are dominated by Ce-4$f$ and O-$2p$ states with smaller contributions from Ce-$4d$ and $5p$ states. See SI Figure 6-9 for orbital-projected density of states. The states at the valence band minima (VBm) are mostly O-2$p$ states. 

Negligible changes occur in the states present at the VBm and CBM upon doping. For all but the Be-doped GBs, $d$, $p$ and $s$ states of the solute atom are present at the CBM, resulting in distorted cubic bonding of the solute-O bonds at the GB, see Figure \ref{fig:Figure3}d and SI Figure 6-9. For Be-doped GBs, only $p$ and $s$ states of the Be are present at the CBM, indicating a strong propensity of Be to form tetrahedral Be-O bonds as shown in Figure \ref{fig:Figure3}c. 
    
\begin{figure}[ht]
        \includegraphics[width=0.5\linewidth]{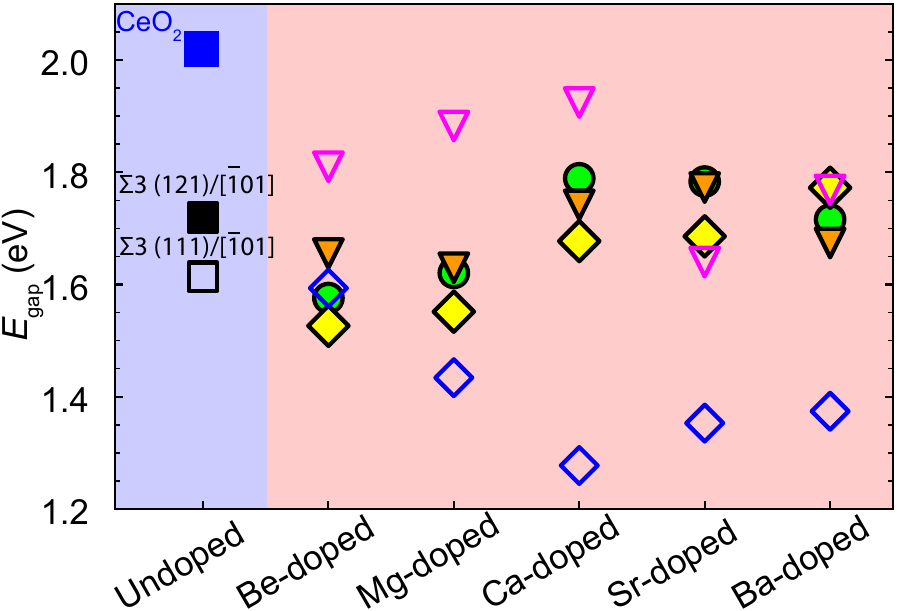}
        \caption{a) The \egap\ for each solute-GB configuration. The undoped GBs are represented with black squares and the bulk ceria by blue square. The \stoot\ GB shows minor site and solute dependence while the \stooo\ GB shows both site and solute dependence.}
        \label{fig:Figure4}
    \end{figure}

 Figure \ref{fig:Figure4} shows that, relative to bulk ceria, the presence of the GBs and solutes has a significant impact on the band gap. The presence of the planar defect, the GB, results in a decrease in the \egap\ with calculated \egap\ values for the \stooo\ and \stoot\ GB of 1.61 eV and 1.72 eV, respectively. The incorporation of solutes can modulate the band gap further, by up to 0.3 eV relative to the undoped GB. A close inspection of the occupied states in the DOS of AEM doped GBs reveals that the states well below the Fermi level alter the Ce-O bonded states in a manner that the \egap\ decreases with respect to the bulk.

The \stooo\ GB shows greater site and solute dependence for the \egap\ than the \stoot\ GB. Although strain has been shown to strongly alter band gap of bulk ceria\cite{Ahn2014,Wen2015}, the increased site sensitivity of the \egap\ for the \stooo\ GB is not correlated with changes in the lattice vectors or volume, see SI Figure 4. Based on this observation we can infer that perhaps the changes in bond strains and the atomic environment have a greater impact for GB cores which are atomically coherent and have bulk like packing densities.
    
For the \stoot\ GBs, the band gap changes are well correlated with the average Ce-O bond distances, see SI Figure 2, and fluctuate around the undoped \stoot\ GB band gap. However, for the doped \stooo\ GB the band gaps appear to be modulated by three main factors: bond strain, local atomic environment of the GB core, and the ionic radii of the solute atom. The coordination-deficient sites (magenta triangle) in the \stooo\ GB have less strain than the fully-coordinated sites (blue diamond) resulting in a linear increase in the band gap, see SI Figure 10. This trend continues until Sr and Ba where the band gap values decrease, an effect which most likely originates from the increased bond strain which extends well into the bulk and can be seen in SI Figure 10. The fully-coordinated sites, blue diamond symbols, are unable to relax since they are sterically hindered by the surrounding anions. The increased strain for these sites which increase the hybridization between the Ce 4$f$ -- O 2$p$ -- M $nd$, $np$ and $ns$ states, where M = Ca, Sr, and Ba and $n$ is the principle quantum number, decreasing the band gap which can be seen in SI Figure 6. For both the GBs, the Ba-doped ceria maintains a similar band gap compared with the undoped GB samples for all sites considered. 
    
In all, the band gap in ceria is considerably affected by the presence of both planar and solute defects. This can result in heterogeneous electronic properties in experimentally synthesized nanocrystalline ceria. The changes in the band gap due to both solutes and the presence of GBs are correlated with the local atomic structure of the GB, average Ce-O bond distance, and the bond strain. Furthermore, the sensitivity of the electronic structure may be modulated by the GB packing density. For close packed GBs, sites which are sterically hindered may have increased hybridization decreasing the band gap, where unhindered sites show a linear increase in the band gap.

\section{Conclusions}
In conclusion, we use DFT with GGA+U functional to examine the structure, stability and electronic properties of undoped and alkali-earth metal doped GBs in ceria. We studied two high-angle grain boundaries, the \stooo\ and the \stoot\ GB and find that the \stooo\ GB is thermodynamically more stable than the \stoot\ GB due to its larger atomic coherency at the GB interface. 

Considering all the substitutional sites in the GB core, we find that when the GBs are doped with $\sim$20\% AEM solutes, the GB energies of ceria will depend strongly on the substitutional site's coordination numbers and its local atomic structure. We identify the lowest energy substitutional sites for each AEM dopant and find that Ca, Sr and Ba solutes stabilize the GBs but Be and Mg solutes render the GBs unstable. The enhancements in the GB stability upon addition of Ca, Sr and Ba can be attributed to similarity in the ionic radii of the solutes and Ce as well as the closely matching coordination of the solute in its native oxide and the ceria lattice. The electronic density of states of doped GBs reveals that no defect states are present in or above the band gap of the AEM doped ceria, which is highly conducive to maintaining low electronic mobility in these ionic conductors. The electronic properties, unlike the thermodynamic stability, exhibit complex inter-dependence on the structure and chemistry of the host and the solutes. The presence of dopants can modulate the band gap of ceria up to 0.3 eV in comparison to the undoped ceria with GBs. 

In the future, advances in computational methods and computing power can enable a comprehensive first-principles based study of more GB structures, solute concentrations as well as the coordinated transport of oxygen-vacancies and ions. Our work serves as a guide to these future studies, making an impact on the design of more efficient oxide based ionic conductors. 

\begin{acknowledgement}
This work was supported by the National Science Foundation (NSF), Department of Energy (DE) and Arizona State University. TB and  PAC are supported by DE SC0004954, NSF DMR-1308085 and DMR-1840841. AS is supported by NSF DMR-1906030. This work used the Extreme Science and Engineering Discovery Environment (XSEDE), which was supported by National Science Foundation grant number TG-DMR150006. The authors acknowledge Research Computing at Arizona State University for providing HPC resources that have contributed to the research results reported within this paper.  

\end{acknowledgement}

\begin{suppinfo}
Supporting information provides a detailed analysis of the GB structure, symmetry, and associated cation site symmetries for each GB to elucidate the connection between local GB structure and strain. The symmetry, structural, and energetic properties of all stable bulk oxide compounds with stoichiometry MO are compiled. The average Ce-O bond distance, $z$-axis GB expansion, and percent change in the volume of the solute ion doped GB structures is also shown. All element and orbital projected DOS and bond strain maps are provided for each solute-GB structure. 
\end{suppinfo}

\providecommand{\latin}[1]{#1}
\makeatletter
\providecommand{\doi}
  {\begingroup\let\do\@makeother\dospecials
  \catcode`\{=1 \catcode`\}=2 \doi@aux}
\providecommand{\doi@aux}[1]{\endgroup\texttt{#1}}
\makeatother
\providecommand*\mcitethebibliography{\thebibliography}
\csname @ifundefined\endcsname{endmcitethebibliography}
  {\let\endmcitethebibliography\endthebibliography}{}

\end{document}